\documentstyle[graphicx,aps,multicol]{revtex}
\newcommand{\lsim}{{ _< \atop ^\sim}}
\newcommand{\gsim}{{ _> \atop ^\sim}}
\begin{document}

\title{Magnon Heat Transport in
(Sr,La)$_{14}$Cu$_{24}$O$_{41}$}
\author{C.~Hess$^{1,2}$, C.~Baumann$^{1,2}$, U.~Ammerahl$^{1}$, B.~B\"{u}chner$^{2,1}$,
F.~Heidrich-Meisner$^3$, W.~Brenig$^3$, A. Revcolevschi$^4$}
\address{$^1$ II. Physikalisches Institut, Universit\"{a}t zu K\"{o}ln, 50937 K\"{o}ln, Germany\\
$^2$ II. Physikalisches Institut, RWTH Aachen, 52056 Aachen,
Germany\\ $^3$ Institut f\"{u}r Theoretische Physik, Technische
Universit\"{a}t Braunschweig, 38106 Braunschweig, Germany\\ $^4$
Laboratoire de Physico-Chimie, Universit\'{e} Paris-Sud, 91405 Orsay,
France}
\date{\today}
\maketitle
\begin{abstract}
We have measured the thermal heat conductivity $\kappa$ of the compounds
Sr$_{14}$Cu$_{24}$O$_{41}$ and Ca$_{9}$La$_5$Cu$_{24}$O$_{41}$ containing
doped and undoped spin ladders, respectively. We find a huge anisotropy of
both, the size and the temperature dependence of $\kappa$ which we
interpret in terms of a very large heat conductivity due to the magnetic
excitations of the one-dimensional spin ladders. This magnon heat
conductivity decreases with increasing hole doping of the ladders. The
magnon heat transport is analyzed theoretically using a simple kinetic
model. From this analysis we determine the spin gap and the temperature
dependent mean free path of the magnons which ranges by several thousand
$\rm \AA$ at low temperature. The relevance of several scattering channels
for the magnon transport is discussed.
\end{abstract}
\begin{multicols}{2}
\narrowtext


Low-dimensional quantum magnets have attracted much attention in  recent

years among both experimental and theoretical physicists. On the one hand
these compounds serve as model systems for a comparison between experiment
and theory, since exact solutions or numerical treatments of the model
Hamiltonians yield clear-cut predictions. On the other hand unusual ground
states and magnetic excitations are present, in particular in
quasi-one-dimensional spin systems. A prominent example is the quantum
disordered spin liquid state with a spin gap which is well established in
frustrated and dimerized spin chains and in spin ladders
\cite{Dagotto96,Rice97,Rice98,Dagotto99,Johnston00a,Johnston00b,Sachdev00,Klumper00}.

Usually the magnetic excitations of these spin systems are experimentally
studied by measuring spectroscopic or thermodynamic quantities which
reveal information on the magnetic ground state and the excitation spectra
as a function of energy and momentum. In principle, dispersive magnetic
excitations should also contribute to a transport property, i.e. the
thermal heat conductivity $\kappa$. The experimental investigation of the
magnon heat transport could give interesting complementary information on
the magnetic excitations, such as dissipation and scattering of magnons,
similar as the study of electronic transport properties does in metals.

Recently, several studies of thermal heat conductivity in low dimensional
spin systems have been performed and magnetic contributions to $\kappa$
are discussed in one-dimensional spin systems such as e.g. CuGeO$_3$
\cite{Ando98,Vasilev98,Takeya00} and Sr$_2$CuO$_3$ \cite{Sologubenko00a} as well as in two dimensional
cuprates~\cite{Nakamura91,Cohn95}. However, in many compounds the
interpretation of the data is controversial, since an unambiguous
discrimination of different contributions to $\kappa$ is difficult or even
impossible\cite{Hofmann01}. Very convincing experimental evidence for a
magnon heat transport has been presented by Sologubenko {\it et al.} for
(Sr,Ca)$_{14}$Cu$_{24}$O$_{41}$~\cite{Sologubenko00}. The crystal
structure of this compound contains two quasi-one-dimensional magnetic
subsystems along the $c$-axis. One subsystem is a sheet like arrangement of
Cu$_2$O$_3$ two-leg ladders, where the Cu $S=\frac{1}{2}$ spins are
strongly coupled via a Cu-O-Cu superexchange, while the other subsystem is
an array of CuO$_2$ $S=\frac{1}{2}$ spin chains with weak magnetic
interactions. While even the stoichiometric compound
Sr$_{14}$Cu$_{24}$O$_{41}$ is hole doped the holes are predominantly
located in the chains \cite{Osafune97,Nucker00}. Changing the composition
i.e., the Sr, Ca, La content in (Sr,Ca,La)$_{14}$Cu$_{24}$O$_{41}$, alters
concentration and distribution of the holes. This results in drastic
changes of the magnetic properties of the chains, whereas a large spin gap
in the ladders of the order of 400$\,$K is observed in all compounds
\cite{Kumagai97,Magishi98,Eccleston98,Katano99}.  However, the charge
transport is determined by the holes in the ladders which are already
present in (Sr,Ca)$_{14}$Cu$_{24}$O$_{41}$. Undoped ladders are only found
in systems containing a large amount of trivalent ions, e.g. La.


In this paper we report on the thermal conductivity of
Sr$_{14}$Cu$_{24}$O$_{41}$ and Ca$_{9}$La$_5$Cu$_{24}$O$_{41}$ parallel
and perpendicular to the chain/ladder direction. In the undoped ladders of
Ca$_{9}$La$_5$Cu$_{24}$O$_{41}$ the magnon contribution is very large and
exceeds the phonon contribution by nearly two orders of magnitude. A
simple approach to describe the energy transport due to magnetic
excitations in the ladders is presented.


We have grown single crystals of Ca$_{14}$La$_5$Cu$_{24}$O$_{41}$
and Sr$_{14}$Cu$_{24}$O$_{41}$ by the traveling solvent floating
zone method \cite{Ammerahl98,Ammerahl99}. Using a standard steady
state method measurements of $\kappa$ have been performed on
pieces cut along the principal axes with a typical length of
2$\,\rm mm$ along the measuring direction and of about 0.5$\,\rm
mm$ for the two other directions. The thermal gradient has been
determined with an Au/Fe-Chromel thermocouple.


Fig.~1 presents $\kappa$ of Sr$_{14}$Cu$_{24}$O$_{41}$ as a function of
temperature $T$ measured along the three crystallographic axes
($\kappa_a$, $\kappa_b$, $\kappa_c$). A striking anisotropy of both,
absolute value and temperature dependence of the thermal conductivity is
apparent. Only for the $b$-axis we find the qualitative behavior expected
for phonon heat transport. At low $T$ the occupation of phonon states
implies an increase of the heat conductivity, whereas $\kappa_b$ decreases
at high $T$ due to increased phonon scattering. $\kappa_a$ deviates
slightly from this usual phonon thermal conductivity of insulators (see
inset of Fig.~1): At high temperature we find a slight increase with
increasing $T$. This small increase of $\kappa_a$ might be related to the
complexity of the phonon spectrum and/or unusual scattering processes
possibly related to the anomalous T dependence of the lattice
constants~\cite{Ammerahl00a}. We note that small deviations from the T
dependence expected for the phonon heat transport as we observe for
$\kappa_a$ in Sr$_{14}$Cu$_{24}$O$_{41}$ are rather common for complex
transition metal oxides. It is still possible and meaningful to discuss
this data in the framework of a phonon heat conductivity $\kappa_{ph}$.
\begin{figure}
\label{fig1}
\includegraphics [width=0.9\columnwidth,clip] {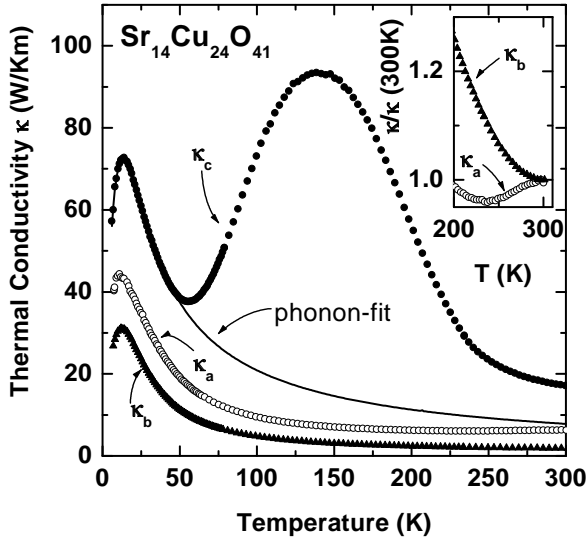}
\caption[]{Anisotropic thermal conductivity $\kappa_a$, $\kappa_b$ and $\kappa_c$ of Sr$_{14}$Cu$_{24}$O$_{41}$ as a
function of temperature. Inset: $\kappa_a$ and $\kappa_b$ normalized to the value at $300\,$K.}
\end{figure}
This is impossible in the case of $\kappa_c$, i.e. for the heat
transport along the chain/ladder direction in
Sr$_{14}$Cu$_{24}$O$_{41}$. In this case the phonon heat
transport can only explain the data below $\approx 40\,$K where
$\kappa_c$ exhibits a low $T$ maximum very similar to $\kappa_a$
and $\kappa_b$. The absolute value is about 1.6 times larger than
for $\kappa_a$. Since the velocity of sound and the elastic
constants do not show a pronounced anisotropy~\cite{Konig97} one
has to attribute the anisotropy of $\kappa_{ph}$ to anisotropic
phonon scattering. Such an anisotropy of the phonon mean free
path is also observed in other transition metal
oxides~\cite{Nakamura91,Hess99} with low dimensional structure
elements.

The most striking observation is the behavior of $\kappa_c$ for
temperatures above $\approx 40\,$K. With increasing $T$ $\kappa_c$
increases strongly and a second very pronounced maximum of the heat
conductivity occurs at $T \simeq 140\,$K. Further increasing the
temperature causes a sharp decrease of $\kappa_c$ and for temperatures
above 250~K the heat conductivity seems to saturate at a still rather
large value of about 18~W/Km. Both the very large values of $\kappa_c$ at
intermediate $T$ and the strange temperature dependence with the
pronounced second maximum differ drastically from the usual $\kappa_{ph}$.
We mention that our data for Sr$_{14}$Cu$_{24}$O$_{41}$ agree
qualitatively with those reported by Sologubenko {\it et al.}
\cite{Sologubenko00}. There are, however, strong deviations to the data of Kudo {\it et al.} \cite{Kudo01}
who report much smaller $\kappa$ for all lattice directions and $T$
probably due to enhanced defect scattering in their samples. Comparing our
data for $\kappa_c$ quantitatively with those in Ref. \cite{Sologubenko00}
reveals a perfect agreement for the low $T$ maximum while the second peak
is about 30$\%$ larger in our sample. At present we are systematically
studying different crystals in order to understand the origin of these
quantitative deviations.

As discussed in Ref.~\cite{Sologubenko00} it is reasonable to attribute the high temperature maximum of $\kappa_c$ to
magnetic excitations propagating along the spin ladders. The electronic heat transport as estimated from resistivity
data and the Wiedemann-Franz-law is negligible. The $T$-dependence, anisotropy of $\kappa$ and the absolute value of
$\kappa_c$ can not be explained by a phonon heat transport and the data for Ca$_{14}$La$_5$Cu$_{24}$O$_{41}$ allow
also to exclude exotic electronic contributions as discussed below.

A determination and analysis of this very unusual magnon contribution
requires a separation of a phonon background. Unfortunately, it is
impossible to determine unambiguously the phonon background for
Sr$_{14}$Cu$_{24}$O$_{41}$. For example, at high $T$ the absolute values
and $T$-dependence are unpredictable, which is clearly demonstrated by the
qualitatively different behavior of $\kappa_a$ and $\kappa_b$. Similar
problems do exist at low $T$, since $\kappa$ is determined by strongly
anisotropic phonon scattering which can not be determined from independent
experimental data or from theory. This means that any determination of the
additional magnetic contribution in Sr$_{14}$Cu$_{24}$O$_{41}$ leads to a
large error for low $T$ and high $T$, i.e. for temperatures where the total
$\kappa$ is not much
larger than $\kappa_{ph}$. In order to estimate the phonon background we
fit the data at low temperatures $T\lsim40\,$K with a Debye-model and
extrapolate the behavior up to 300~K. The result of this description of
$\kappa_{ph}$ is indicated by a solid line in Fig.~1 which is much smaller
than the measured $\kappa$ in the entire temperature range above 50~K.


However,  considering only the data for Sr$_{14}$Cu$_{24}$O$_{41}$ one can question any non-phononic heat conductivity
at $T \gsim 250$~K. A much more reliable separation of phonon and magnon heat transport is possible in the case of
Ca$_{9}$La$_5$Cu$_{24}$O$_{41}$ whose thermal conductivities along the $a$- and $c$-axes are shown in Fig.~2. At low
$T$ the thermal heat conductivity of this compound which contains undoped spin ladders and only slightly doped spin
chains~\cite{Osafune97,Nucker00} is much smaller than in stoichiometric Sr$_{14}$Cu$_{24}$O$_{41}$ for both
directions. It is straightforward to attribute this isotropic suppression of $\kappa_{ph}$ at low $T$ to structural
defects. On the one hand a lattice site is occupied by randomly distributed $\rm Ca^{2+}$ and $\rm La^{3+}$ ions with
strongly different sizes which is known to suppress $\kappa_{ph}$. On the other hand x-ray diffraction studies show
structurally disordered chains in Ca$_{9}$La$_5$Cu$_{24}$O$_{41}$. Surprisingly,  $\kappa_c$ strongly increases above
$40\,$K  in spite of the structural disorder. At intermediate and high $T$ $\kappa_c$ is even larger than in
stoichiometric Sr$_{14}$Cu$_{24}$O$_{41}$ and the room temperature value is comparable to that found in metals. In
contrast to that a very small heat conductivity is found along the $a$-axis in the entire temperature range. Reduced
size and $T$ dependence are typical for $\kappa_{ph}$ of a compound with many structural defects. It is apparent from
Fig.~2 that this strong damping of $\kappa_{ph}$ which is inferred from $\kappa_a$ as well as from the low T behavior
of $\kappa_c$ enables a reliable determination of the additional contributions to $\kappa_c$ above $\approx 40$~K.
\begin{figure}
\label{fig2}
\includegraphics [width=0.9\columnwidth,clip]{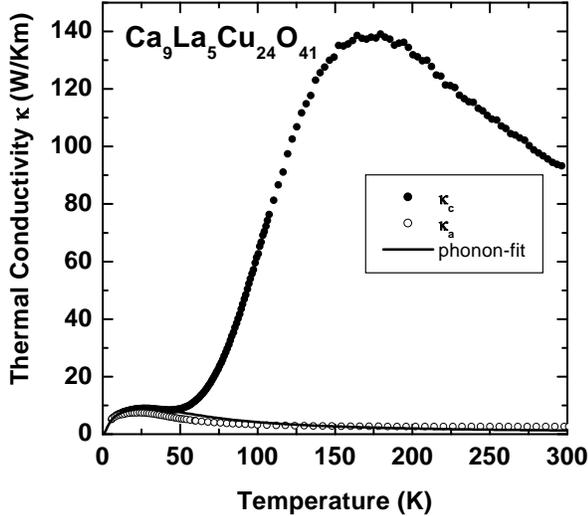}
\caption[]{Thermal conductivities of Ca$_{9}$La$_5$Cu$_{24}$O$_{41}$
as a function of temperature measured along the $a$- and $c$-axes,
$\kappa_a$ and $\kappa_c$, respectively. The solid line represents
an estimate of the phonon contribution to $\kappa_c$.}
\end{figure}

Before we turn to a quantitative analysis of $\kappa_{mag}$ we mention
several conclusions which can be drawn from the qualitative behavior of
$\kappa$ in Ca$_{9}$La$_5$Cu$_{24}$O$_{41}$ and, in particular, from the
comparison to the findings in the stoichiometric compound. The suppression
of $\kappa_{ph}$ which coincides with an enhancement of $\kappa_c$ at high
$T$ gives very strong evidence that there are two independent

contributions to the heat transport. Moreover, some thinkable origins of
the large additional contributions to $\kappa_c$ in
Sr$_{14}$Cu$_{24}$O$_{41}$ can be ruled out from the additional data.
Excitations of the chains are irrelevant, since the magnetic and
electronic properties of the heavily doped chains in
Sr$_{14}$Cu$_{24}$O$_{41}$ showing charge order, spin gap and dimerization
are completely different from the long range ordered chains in
Ca$_{9}$La$_5$Cu$_{24}$O$_{41}$~\cite{Matsuda98,Ammerahl00b}. Heat
transport due to collective electronic excitations as for example
suggested for the thermal conductivity of the charge density wave compound
K$_{0.3}$MoO$_3$~\cite{Bihar97} could be relevant in the strongly doped
charge ordering Sr$_{14}$Cu$_{24}$O$_{41}$ whereas it is certainly
irrelevant in Ca$_{9}$La$_5$Cu$_{24}$O$_{41}$ due to the strongly reduced
hole content. Summarizing these arguments we have to state that the
findings in Ca$_{9}$La$_5$Cu$_{24}$O$_{41}$ strongly support the
interpretation of the additional contribution to $\kappa_c$ in terms of a
magnon heat transport along the spin ladders. This magnon heat
conductivity $\kappa_{mag}$ is apparently not suppressed by structural
disorder and, moreover, a smaller hole doping seems to enhance
$\kappa_{mag}$ at high $T$.

\begin{figure}
\label{fig3}
\includegraphics [width=0.9\columnwidth,clip]{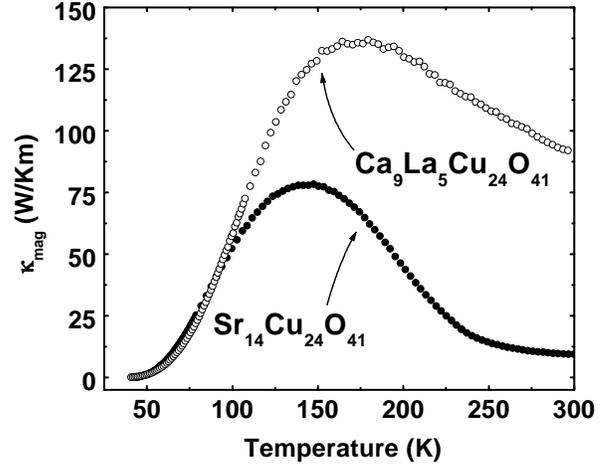}
\caption[]{Magnon thermal conductivities of Ca$_{9}$La$_5$Cu$_{24}$O$_{41}$ and  Sr$_{14}$Cu$_{24}$O$_{41}$ as a
function of temperature.}
\end{figure}

In Fig.~3 we show $\kappa_{mag}$ for Sr$_{14}$Cu$_{24}$O$_{41}$ and
Ca$_{9}$La$_5$Cu$_{24}$O$_{41}$ which are derived by subtracting the
Debye-fits of the phonon contribution from the measured $\kappa_c$. For $T
\lsim100$~K absolute value and temperature dependencies of $\kappa_{mag}$ are similar in the two compounds.
However, pronounced differences occur at higher $T$. The magnetic heat
conductivity on Sr$_{14}$Cu$_{24}$O$_{41}$ $\kappa_{mag}$ strongly
decreases above $\approx 150\,$K. This decrease is much less pronounced
and is found at higher temperature (above $\sim 200$~K) in
Ca$_{9}$La$_5$Cu$_{24}$O$_{41}$ where we find a very large $\kappa_{mag}
\simeq 100$~W/Km even at room temperature.


In order to analyze the magnon heat transport theoretically we start from
the simple kinetic expression
\begin{equation}\label{model}
\kappa_{mag} = \frac{d}{dT}\sum_k v_k \epsilon_k n_k l_k
\end{equation}
where $v_k$, $\epsilon_k$, and $l_k$ denote velocity, energy and mean free
path of the magnetic excitations.  Assuming a momentum independent mean free
path, i.e. $l_k \equiv l_{mag}$ and using the distribution function
\begin{equation}\label{distrib}
n_k=\frac{3}{3+e^{\frac{\epsilon_k}{k_BT}}}
\end{equation}
for the triplet excitations yields the following expression for the magnon
heat conductivity:
\begin{equation}\label{kappamag}
\kappa_{mag}=\frac{3 N l_{mag}}{\pi\hbar k_BT^2}\int_{\Delta_{ladder}}^{\epsilon_{max}}
\frac{exp(\epsilon/k_BT)}{(exp(\epsilon/k_BT)+3)^2}\epsilon^2d\epsilon~.
\end{equation}
Here, $N$ is number of ladders per unit area and $\epsilon_{max}$ is the band maximum of the spin excitations which is
at approximately 200 meV in (Sr,Ca,La)$_{14}$Cu$_{24}$O$_{41}$. Note,  that the distribution function (\ref{distrib})
is different from a Bose distribution. This difference is used to account, on average, for the hard-core constraint of
no on-site double-occupancy for the triplet excitations. While such a form of the distributions function would
describe the occupation of  a {\em local} triplet excitation exactly, it is only meant as an approximate phenomenology
regarding the momentum space distribution - which is unknown for the present case. The primary motivation for of this
type of a triplet distribution is to suppress unphysically large triplet densities at higher temperatures. From
Eq.~\ref{kappamag}  it is apparent that the particular form of the magnon dispersion does not enter the heat
conductivity of a one-dimensional triplet gas within a kinetic description - an effect which has not been noticed in
other studies~\cite{Sologubenko00}. In particular, the momentum or energy dependence of the magnon velocity $v$ which
emerges from Eq.~\ref{model} plays no r\'ole in our approach.  Note that the assumption of a $k$-independent  mean free
path implies a $k$-dependent scattering rate since $v$ depends on the wave number. While this assumption seems
reasonable for defect scattering more involved scenarios will apply in particular to inelastic scattering. Within our
treatment the magnon heat

conductivity for experimentally relevant temperatures $T \ll \epsilon_{max}/k_B$ mainly depends on two parameters: the
spin gap $\Delta_{ladder}$ and the mean free path $l_{mag}$. We mention, that (\ref{kappamag}) differs from an
expression used by Sologubenko {\em et al.}~\cite{Sologubenko00} for the heat-conductivity of one-dimensional bosons
not only by the distribution function~(\ref{distrib}) but also by an overall factor of three accounting for the
triplet degeneracy.


For temperatures below 300~K $\kappa_{mag}$ does not depend significantly
on $\epsilon_{max}$ and in the following analysis of the data in the
framework of Eq.~\ref{kappamag} we use $\epsilon_{max} \simeq 200$~meV
taken from the literature~\cite{Eccleston98,Matsuda00}. It is impossible,
to determine reasonable values of the two remaining parameters $l_{mag}$
and $\Delta_{ladder}$ by fitting the data with the expression in
Eq.~\ref{kappamag} without further assumptions. Therefore we ignore
possible temperature dependencies of the spin gap and, moreover, assume a
constant, temperature independent mean free path at low $T$, i.e. for a
small number of thermally activated phonons and magnons. Knowing the spin
gap from this first step of the analysis we can directly extract the
temperature dependent $l_{mag}$ from the data.

Our assumption of a constant $l_{mag}$ at low $T$ implies that the heat
conductivity is determined by the temperature dependent activation of
magnons in this temperature range. This is in agreement with the
experimental data which roughly follow a simple activated behavior as
displayed in Fig.~4. Note that Eq.~\ref{kappamag} predicts deviations from
the simple activation law and taking into account these corrections
slightly improves the description of the data as demonstrated in Fig.~4.
The spin gaps of $\Delta_{ladder} = 396~$K and $\Delta_{ladder} =
430$~K we obtain are in fair agreement with the results from neutron scattering.
We conclude that both, the temperature dependence of $\kappa_{mag}$ and the
absolute values  of the spin gaps extracted from the data confirm our
analysis. Comparing the left and right columns of Fig.~4 reveals that the
agreement between data and theory is much better for
Ca$_{9}$La$_5$Cu$_{24}$O$_{41}$ than for Sr$_{14}$Cu$_{24}$O$_{41}$. In
the latter case a fair agreement with the theory at constant $l_{mag}$ can
only be obtained in a limited temperature range $55 {\rm K} \lsim T \lsim
85$~K. The deviations at higher temperatures, i.e. for $T \gsim 85$~K
signal a significant temperature dependence of the mean free path, while
the differences at low temperatures $T \lsim 55$~K are most probably due
to the uncertainty of the phonon heat conductivity which is much larger
than $\kappa_{mag}$ in this temperature range (see Fig.~1).  The
separation of $\kappa_{mag}$ and $\kappa_{ph}$ is much more reliable in
Ca$_{9}$La$_5$Cu$_{24}$O$_{41}$ and, indeed, the thermal conductivity for
$T \lsim 100$~K is well described by our model with constant $l_{mag}$.

\begin{figure}
\label{fig4}
\includegraphics [width=.9\columnwidth,clip]{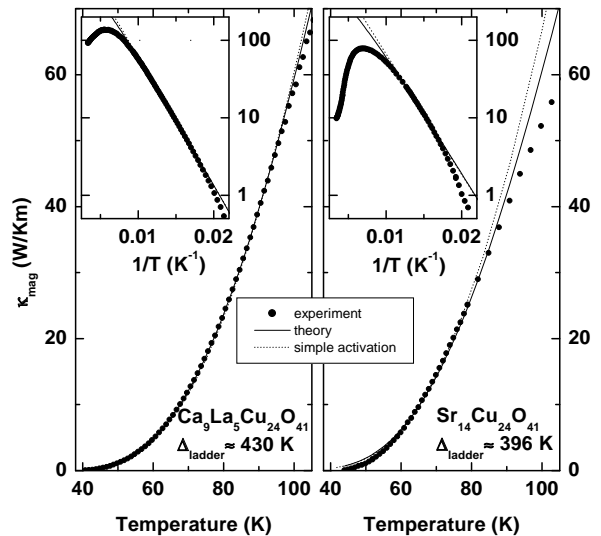}
\caption[]{Temperature dependence of the magnon thermal
conductivities of Ca$_{9}$La$_5$Cu$_{24}$O$_{41}$ (left)
and Sr$_{14}$Cu$_{24}$O$_{41}$ (right) at low temperatures in comparison
with fit results to the data (Dotted lines: simple activation, solid
lines: 1d-model for $\kappa$). Insets: Arrhenius representation of $\kappa$
and fit results.}
\end{figure}


Based on the preceding analysis of the low temperature behavior of $\kappa_{mag}$ it is possible to determine the
temperature dependence of the mean free path $l_{mag}$ (see Fig.~5). At low temperatures we find a very large
$l_{mag}$ of several thousand {$\rm \AA$}. Within the experimental uncertainty due to the phonon background we can not
resolve differences of the low $T$ saturation values of $l_{mag}$ for the two compounds containing spin ladders with
different hole content. There are, however, unambiguous and drastic differences at higher temperatures. In
Sr$_{14}$Cu$_{24}$O$_{41}$ which contains spin ladders with a rather large hole doping the magnon mean free path is
much smaller and we find a rather strong temperature dependence between 85~K and 220~K. According to our analysis
$l_{mag}$ saturates for higher temperatures and at $T=300$~K the mean free path is about 60 {\rm \AA}. While the
saturation of $l_{mag}$ above 220~K is clearly extractable from our data,  the absolute values of $\kappa_{mag}$ and
$l_{mag}$ strongly depend on the choice of the background, i.e. the extrapolation of $\kappa_{ph}$, and therefore a
much smaller $l_{mag}$ at 300~K is also compatible with the raw data in Sr$_{14}$Cu$_{24}$O$_{41}$. For the second
crystal with nearly undoped spin ladders a  very large $l_{mag}$ at room temperature is out of question. Our analysis
yields a magnetic mean free path larger than $500$ ${\rm\AA}$ at 300~K. Moreover, the overall temperature dependent
decrease of $l_{mag}$ is less pronounced and deviations from the constant low $T$ value start at higher temperatures
than in Sr$_{14}$Cu$_{24}$O$_{41}$.


Our results for $l_{mag}$ in Sr$_{14}$Cu$_{24}$O$_{41}$ are in qualitative agreement with the findings reported by
Sologubenko et al.~\cite{Sologubenko00}. There are, however, quantitative discrepancies due to the different
theoretical description: for a given $\kappa_{mag}$ our analysis leads to a smaller value of $l_{mag}$, mainly due to
the more realistic distribution function for the magnetic excitations which we use in our approach
(Eq.~\ref{distrib}). Yet, the mean free paths which we extract from the analysis of our data are very large. In
Ref.~\cite{Sologubenko00} it was speculated that holes are the main scatterers in the doped spin ladders of
(Sr,Ca)$_{14}$Cu$_{24}$O$_{41}$. This speculation is only partially supported by our results on a crystal with undoped
spin ladders. At low $T$ our data do not show a significant difference between the two crystals indicating that the
very large $l_{mag}$ at low temperature does not depend strongly on the hole content of the ladders. It is likely that
the maximum magnon mean free path is determined by defects. Note however, that one has to discriminate between crystal
defects and ''magnetic defects'', since there is no correlation between the maximum $l_{mag}$ and the maximum phonon
mean free path. The latter is much smaller in Ca$_{9}$La$_5$Cu$_{24}$O$_{41}$ which is obvious from the damping of
$\kappa_{ph}$ at low $T$.
\begin{figure}
\label{fig5}
\includegraphics [width=0.9\columnwidth,clip]{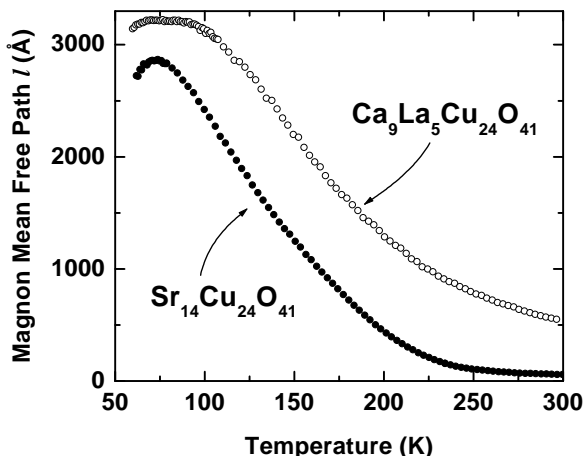}
\caption[]{Magnon mean free paths of
Ca$_{9}$La$_5$Cu$_{24}$O$_{41}$ (open symbols) and
Sr$_{14}$Cu$_{24}$O$_{41}$ (full symbols) as a function of temperature.}
\end{figure}

In contrast to the results at low $T$ we do find a pronounced doping
dependence of the magnon scattering at high $T$. The magnon mean free path
is much smaller for the doped spin ladders of Sr$_{14}$Cu$_{24}$O$_{41}$
and of the order of the  mean hole distance in the ladders
($\approx25\,\rm\AA$) at 300~K \cite{Nucker00}. In the same crystal
resistivity measurements show a pronounced change of their temperature
dependence at $T \simeq 220$~K. This anomaly which is frequently
interpreted in terms of charge ordering is absent in
Ca$_{9}$La$_5$Cu$_{24}$O$_{41}$. Thus, one might argue that  instead of
the hole content, the hole mobility is crucial for the scattering of magnons,
in order to explain the weak doping dependence of $l_{mag}$ at low $T$ and
the strong differences at high $T$. Yet, holes and their mobility are not
the only source of temperature dependent magnon scattering as demonstrated
by our data for the crystal with undoped spin ladders. In this crystal we
find a less pronounced, but qualitatively similar decrease of $l_{mag}$
for $T > 100$~K. At room temperature the number of both, thermally excited
magnons and phonons is already rather large. For example, the inverse
density of magnons at 300~K as calculated for $\Delta_{ladder} = 430$~K
can be interpreted as the "mean distance" of magnons and is already one
order of magnitude smaller than $l_{mag}$.  Further experimental and
theoretical studies are necessary in order to determine the importance of
magnon-defect, magnon-phonon and magnon-magnon scattering.


In summary we have measured the thermal conductivity of hole doped and
undoped spin ladders realized in  the compounds
Sr$_{14}$Cu$_{24}$O$_{41}$ and Ca$_{9}$La$_5$Cu$_{24}$O$_{41}$.
In both cases we find a huge contribution to $\kappa$ due to
magnon heat transport. We have applied a simple kinetic approach to
describe this contributions and have obtained the spin gaps of the ladders and
the temperature dependence of the mean free path from our analysis. Obviously,
at low $T$ scattering of magnons on holes is rather ineffective since
in this temperature region $\kappa_{mag}$ of doped ladders is almost
identical to $\kappa_{mag}$ of the undoped ladders. For both we find a
very large mean free path of magnons of several thousand $\rm\AA$. At high
$T$ increased mobility of holes causes a strong damping of magnon heat
transport in the hole doped ladders due to increased magnon-hole
scattering whereas $\kappa_{mag}$ in the undoped ladders only decreases
slightly. This is reflected in the mean free path which in the former case
at $300\,$K reduces to $l_{mag}\approx60\rm\AA$ which is in the same order
of magnitude than the mean hole distance ($\approx25\rm\AA$). For undoped
ladders $l_{mag}\approx500\rm\AA$ at $300\,$K. This large value is one
order of magnitude larger than the inverse density of magnons at this
temperature and might indicate only weak magnon-magnon scattering effects.

This work was supported in part by the Deutsche Forschungsgemeinschaft through SFB 341 and SP 1073 as well as under
Grant No. BR 1084/1-1 and BR 1084/1-2.

\end{multicols}
\end{document}